\documentclass[aps,prl,twocolumn,superscriptaddress,showpacs]{revtex4}
\usepackage{bm}
\usepackage{times}
\usepackage{amssymb}
\usepackage[dvips]{graphicx}

\begin{document}

\author{A. A. Reynoso}
\affiliation{Instituto Balseiro, Centro At\'omico Bariloche, R\'{\i}o 
Negro, Argentina.}
\affiliation{Consejo Nacional de Investigaciones Cient\'{\i}ficas y T\'ecnicas (CONICET), Argentina}
\author{Gonzalo Usaj}
\affiliation{Instituto Balseiro, Centro At\'omico Bariloche, R\'{\i}o 
Negro, Argentina.}
\affiliation{Consejo Nacional de Investigaciones Cient\'{\i}ficas y T\'ecnicas (CONICET), Argentina}
\author{C. A. Balseiro}
\affiliation{Instituto Balseiro, Centro At\'omico Bariloche, R\'{\i}o 
Negro, Argentina.}
\affiliation{Consejo Nacional de Investigaciones Cient\'{\i}ficas y T\'ecnicas (CONICET), Argentina}
\author{D. Feinberg}
\affiliation{Institut NEEL, CNRS and Universit\'e Joseph Fourier, 
Boite Postale 166,
38042 Grenoble, France.}
\author{M. Avignon}
\affiliation{Institut NEEL, CNRS and Universit\'e Joseph Fourier, 
Boite Postale 166,
38042 Grenoble, France.}
\title{Anomalous Josephson Current in Junctions with Spin-Polarizing
Quantum Point Contacts}

\date{2 May 2008}

\begin{abstract}
We consider a ballistic Josephson junction with a quantum point 
contact in a two-dimensional electron gas with Rashba spin-orbit 
coupling. The point contact acts
as a spin filter when embedded in a circuit with normal electrodes. 
We show that
with an in-plane external magnetic field an anomalous supercurrent 
appears even for zero
phase difference between the superconducting electrodes. In addition, the
external field induces large critical current asymmetries between the two flow
directions, leading to supercurrent rectifying effects.
\end{abstract}
\pacs{74.45.+c, 71.70.Ej, 72.25.Dc, 74.50.+r}

\maketitle

Josephson junctions (JJ) are the basic building blocks for superconducting
electronics with applications that range from SQUID magnetometers to possible
quantum computing devices. In superconductor-normal metal-superconductor
(S-N-S) junctions the supercurrent flow is due to the Andreev states---a coherent superposition of electron and holes states.
These states depend on the electronic structure of the normal material and on the properties of the S-N interface
\cite{golubov,liharev,beenakker1,furu1}. Modern technologies based on two
dimensional electron gases (2DEGs) \cite{takaya,2DEG} or 
nanowires \cite{nanowire} allow for a precise control of such 
electronic properties, and thus of the JJ
characteristics. Moreover, spin-orbit (SO) effects offer
new alternatives to control the spin and charge transport
\cite{winkler,Spintronicsbook}.

Superconducting rectifiers are among the new devices proposed and studied
during the last few years. Most of these proposals are based on the dynamics
of vortices \cite{Zapata,Carapella}. Here we show that in systems with SO-coupling rectifying
properties can be obtained by controlling the spin of the Andreev states. To this end we consider
a ballistic JJ with a quantum point contact (QPC) in a
2DEG with SO interaction. The QPC can be tuned to control the number of transmitting channels and thus the critical current of
the junction \cite{beenakker1,furu1,takaya,muller,kuhn}. On the other 
hand, the QPC with SO coupling may act as a spin filter
producing spin-polarized currents when embedded in a circuit with
normal leads \cite{eto,silvestrov,reynoso2}. The normal current also 
generates an in-plane magnetization---perpendicular to the current
---as well as out-of-plane spin-Hall textures \cite{usaj}. Both effects are
maximized at the core of the QPC \cite{reynoso1}. As the SO-coupling preserves
time-reversal symmetry (TRS), we expect that these peculiarities of the
transmitting channels do not harm the Josephson effect when the leads become
superconducting. However, the Josephson current itself breaks the TRS 
and, as we
show below, it reveals striking effects of the SO-coupling. For example, the
supercurrent generates spin polarization in the 2DEG  \cite{malshukov} and the QPC in a similar way normal
current does  \cite{usaj,reynoso1}. This is due to the distinctive spin texture of
each Andreev state that contributes to the local magnetization in a 
supercurrent-carrying state.

\begin{figure}[b]
\includegraphics[width=.45\textwidth,clip]{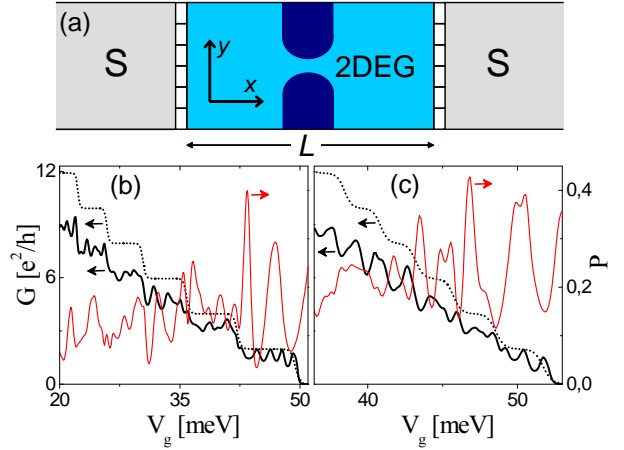}
\caption{(a) Schematic view of the junction. Panels (b) and (c) correspond to
QPC1 and QPC2 simulated with gates placed at $z\!=\!30$ and $90$ nm 
on top of the
2DEG, respectively. The conductance with ideal 2DEG electrodes (dotted line),
with the metallic electrodes described in the text (thick line) and the current
polarization $P$ (thin line) are shown. The Rashba coupling strength is
  $\alpha\!=\!20$ meVnm. }
\end{figure}

More striking effects take place if an external in-plane magnetic
field is applied. Its effect on the supercurrent characteristics depends
on the nature of the junction. In the absence of SO-coupling, the Zeeman field
may generate $\pi$-junctions resembling the case of S-ferromagnet-S junctions
\cite{buzdin,golubov}. For systems with SO-coupling, the existing theories
include the description of perfectly contacted 2DEG junctions \cite{bezuglyi},
wide junctions \cite{dimitrova}, 1D-conductors \cite{krive} and junctions with
quantum dots \cite{dellanna,beenaker2}. In our case, the QPC, the internal
SO field and the external Zeeman field conspire to reveal novel 
effects. Remarkably, we find a critical current $I_c$ that depends on the current flow 
direction. With more than one transmitting channel, the QPC can be tuned to show either a large $I_c$
asymmetry or a perfect symmetry. In this regime the JJ can act as a supercurrent rectifier  \cite{Zapata,Carapella}.
At the origin of this effect is the  anomalous supercurrent---proportional 
to the external field---that appears even for zero phase difference $\phi$ between the two 
superconducting leads \cite{krive,buzdin}.
 Devices
based on InAs-related materials, that present strong gate-tunable Rashba
SO-coupling are good candidates to look for these effects
\cite{takaya,2DEG}. In what follows we present the main results
of the theory.

The total Hamiltonian of the system reads
\begin{equation}
H\!=\!H_{QPC}+H_{R}+H_{L}+H_{C},
\end{equation}
here $H_{QPC}$ describes the central 2DEG with the QPC (see Fig.1(a)). In the effective mass approximation $%
H_{QPC}\!=\!(p_{x}^{2}+p_{y}^{2})/2m^{\ast }+\alpha /\hbar (p_{y}\sigma
_{x}-p_{x}\sigma _{y})+V(x,y)-g\mu _{B}\mathbf{\vec{\sigma} \cdot 
B}$, where the
first two terms are the kinetic energy and the Rashba SO-coupling, 
respectively,
and $V(x,y)$ is the confinement potential that defines the QPC. We 
use a potential that simulates the effect of two
electrodes held at a distance $z$ from the 2DEG \cite{Ferrybook}, 
with a gate voltage controlling the height, $V_g$, of the
potential barrier at the center of the QPC. The
last term in $H_{QPC}$ is the Zeeman energy. The Hamiltonians $H_{R}$ and $%
H_{L}$ describe the right and left superconducting electrodes with an order
parameter $\Delta_{R/L}\!=\!\Delta _{0}e^{\pm i\phi /2}$. Finally, $H_{C}$
describes the contact between the superconductors and the 
2DEG.

For the numerical calculations we discretized
the space, mapping the Hamiltonian (1) onto a tight-binding-like
model. We use a square lattice with hopping matrix elements $t_{N}$ and $%
t_{S}$ for the normal (2DEG) and superconducting materials, respectively \cite{yeyati}. 
The microscopic parameters of the normal region
correspond to InAs-like materials \cite{takaya}: the effective
mass is $m^*\!=\!0.045m_e$ and the electron density 
$n\sim10^{12}cm^{-2}$ (the Fermi energy is $E_F\sim 53$ meV). We take the total
length of the junction $L\!=\!1.2 \mu$m and analyze two point contacts denoted
as QPC1 and QPC2 corresponding to different values of $z$. We use $\Delta_{0}\!=\!1.5$ meV and the 
coherent length
$\xi_{0}\!=\!\hbar v_{F}^{s}/\Delta_{0}\!=\!43$ nm corresponding to Nb films,
$v_{F}^{s}$ is the Fermi velocity of the superconductor \cite{takaya,2DEG}.

We evaluate the normal and anomalous propagators that
contain the information of all the physical quantities of interest. For $%
\Delta _{0}\!=\!0$ the normal conductance $G\!=\!\sum_{\sigma ,\sigma ^{\prime
}}G_{\sigma ,\sigma ^{\prime }}$ is evaluated using the conventional
Landauer-like formulation \cite{Ferrybook}. Here $G_{\sigma
,\sigma ^{\prime }}$ is the contribution to the conductance due to incident
electrons with spin $\sigma$ that are transmitted with spin $\sigma^{\prime}$.
The spin polarization of the current is defined as $P\!=\!$ $\sum_{\sigma
}(G_{\sigma ,\uparrow }\!-\!G_{\sigma ,\downarrow })/G$. These quantities
characterize the QPC in the normal state. For $\Delta _{0}\neq 0$ we calculate the Josephson current flowing 
through the right N-S interface, \cite{yeyati}
\begin{equation}
I(\phi )\!=\!\mathrm{i} \frac{e}{\hbar }\sum_{i\in N,j\in
S}[t_{SN}^{i,j}\left\langle \psi ^{\dag }(x_{j})\psi (x_{i})\right\rangle
\!-\!t_{SN}^{j,i}\left\langle \psi ^{\dag }(x_{i})\psi (x_{j})\right\rangle ]
\end{equation}
here $x_{i}$ and $x_{j}$ are coordinates at the edge of the 2DEG and the
superconducting electrode, respectively, $t_{SN}^{i,j}$ is the hopping matrix
element connecting neighboring sites at the interface and the field operator
$\psi ^{\dag }(x)\!=\!(\psi _{\uparrow }^{\dag }(x),\psi _{\downarrow 
}^{\dag }(x))$
creates an electron at coordinate $x$. We choose $ t_{SN}^{i,j} 
\!=\!(t_{N}+t_{S})/2$, decreasing this value increases the
normal scattering at the interface producing narrow resonances within the
central 2DEG region with strong influence on the Josephson current \cite{kuhn}.

\textit{The zero-field case}: The normal conductance of the system 
presents clear
structures on top of the plateaus, shown in Fig. 1(b) and 1(c) as 
broad resonances originated in the scattering at the electrode-2DEG
interfaces. As shown in the same figure these QPCs generate spin-polarized
currents with a polarization $P$ in the range $[0$-$0.6]$ depending on the
strength of the SO-coupling and the number of transmitting channels.

\begin{figure}[t]
\includegraphics[height=7cm]{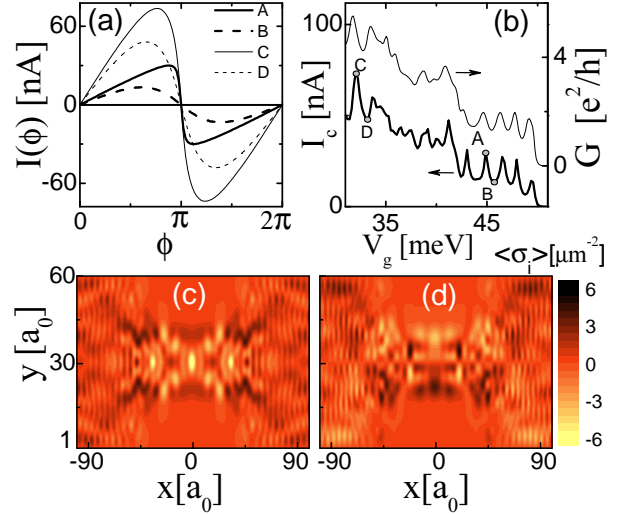}
\caption{(a) Current-phase relation (CPR) for different gate voltages in QPC1.
(b) Critical current and conductance as function of $V_g$. The points labelled
with letters indicate the parameters of the curves of panel (a). In (c) and (d)
color maps of the y and z magnetization calculated with the critical current of
curve C of panel (a) are shown. The SO strength is $\alpha\!=\!20$ 
meV.nm and the lattice parameter is $\mathrm{a}_0=3$nm.}
\label{FIG2}
\end{figure}

For superconducting contacts, the current-phase relation (CPR) is shown in Fig.
2 for different values of the parameters. We observe two characteristic CPR:
resonant-like and tunnelling-like (sinusoidal) relations. As $V_{g}$ 
changes the
junction alternates between these two behaviors. The critical current 
$I_{c}$ is
defined as the maximum current in the CPR. The dependence of $I_{c}$ and the
conductance of the normal state $G$\ on $V_{g}$ is shown in Fig. 2(b). The
structure of both curves is similar with the peaks located at the same position
showing that the maximums of $I_{c}$\ are due to one-electron resonances
\cite{furu1}. The structure of the Andreev spectrum includes a number of
(dispersive) states with phase-dependent energies and (non-dispersive) states
confined at each side of the constriction, details will be presented elsewhere
\cite{reynoso3}. When there is only one transmitting channel, the Josephson
current is dominated by a single (spin) pair of dispersive Andreev states. As
$V_g$ decreases and the QPC opens, the current is the superposition of
contributions due to different pairs of states.

Due to the SO-coupling each Andreev state has a well defined spin texture which
in the presence of a supercurrent contributes to the local
magnetization. As an illustration we calculate the magnetization
components $\left\langle S_{y}(x_{i})\right\rangle$ and $\left\langle
S_{z}(x_{i})\right\rangle $ for all the lattice sites within the 2DEG. As shown
in Fig. 2(c) and 2(d), the supercurrent-induced steady magnetization has an
in-plane component perpendicular to the current direction and an out-of-plane
component with the spin-Hall structure \cite{malshukov,usaj}.

\textit{Effect of external in-plane magnetic fields}: The spin 
texture of each Andreev
state has a component along the $y$-direction. In general, for a 
spin-polarizing
QPCs, a pair of dispersive Andreev states---indicated by $\left\vert
+\right\rangle $ and $\left\vert -\right\rangle $---have $\left\langle
+|S_{y}|+\right\rangle \neq -\left\langle -|S_{y}|-\right\rangle $.
Consequently, with an external field $B$ in the $y$-direction the 
absolute value
of their Zeeman shifts is different. The effect is illustrated in Fig. 3(a) and
3(b). In QPCs with a single transmitting channel we follow a resonance as the
SO-coupling $ \alpha$ is increased.
\begin{figure}[t]
\includegraphics[width=8cm]{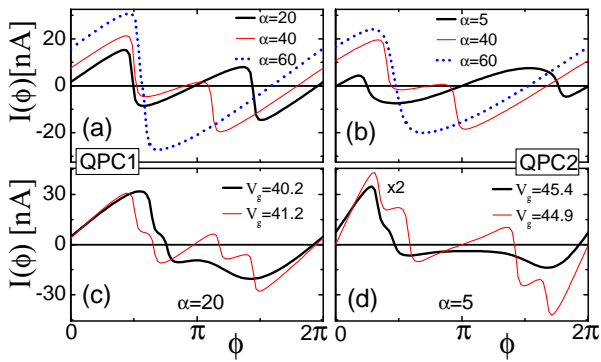}
\caption{CPR for QPC1 (panels (a) and (c)) and QPC2 (panels (b) and 
(d)) with an
external field of $g \mu_B B\!=\!0.3$ meV. In (a) and (b) the QPC 
height $V_g$ is
set to follow a resonance at the first conductance plateau for different values
of the SO coupling $\alpha$ (in meVnm). In (c) and (d) the $I(\phi)$
for different $V_g$ (in meV) at the second and third plateaus are
shown. Note the asymmetry between $I(\phi)>0$ and $I(\phi)<0$.} \label{FIG3}
\end{figure}
For small $\alpha$ we observe the characteristic behavior of a 
resonant state in
the presence of a field. The distance between the step-like changes of the
current is a measure of the Zeeman splitting. As $\alpha$ increases the Zeeman
shift of one of the Andreev states decreases and changes sign (the step-like
structure crosses the $\phi\!=\!\pi$ point). In the large $\alpha$ limit
$\left\langle +|S_{y}|+\right\rangle \approx \left\langle 
-|S_{y}|-\right\rangle
$ and the CPR shows a single step. In this limit, the QPC acts as a 
very efficient spin filter in the first
plateau \cite{eto,reynoso1}: all transmitted electrons have 
essentially the same
spin orientation. This results in an anomalous supercurrent for $\phi\!=\!0$.
To estimate it, we may assume a smooth QPC and evaluate the
phase shift $\vartheta _{n}(E)$ acquired by an electron travelling from one
superconducting electrode to the other in the WKB approximation. In $%
\vartheta _{n}(E)$ the index $n\!=\!\pm $ is the channel index and $E$ is the
energy measured from the Fermi energy. For small $E$ we
have
\begin{equation}
\vartheta _{n}(E)\approx k_{F}\lambda _{n}+\frac{E}{\hbar v_{F}}\delta _{n},
\end{equation}
here $k_{F}$ ($v_{F}$) is the Fermi wavevector (velocity) in the 2DEG,
$\lambda_{n}\lesssim L$ and $\delta_{n}\gtrsim L$ are related to the effective
length of the junction at the Fermi energy. Assuming that at resonance the
scattering at SN interface plays no important role we have \cite{krive}
\begin{equation}
I(\phi )\!=\!\frac{ev_{F}}{\pi }\sum_{n}\frac{1}{\delta _{n}}\Omega (\phi +\mu
_{n}\frac{B}{\Delta _{n}}),
\end{equation}%
here $\Omega (x)$ is a periodic function with $\Omega (x)\!=\!x$ for 
$|x|<\pi $,
$\Delta _{n}\!=\!\hbar v_{F}/\delta _{n}$ and $\mu _{n}B$ is the 
Zeeman shift. Then,
in Fig. 3(a) and 3(b) the step-like structures correspond to $\phi 
\!=\!\pi -\mu
_{n}B/\Delta _{n}$ and the anomalous current 
$I(0)\!=\!(2e/h)\sum_{n}\mu _{n}B$ is
given by the total Zeeman energy of the transmitting channels---note 
it does not
depend on $L$.
Non-linear effects with larger anomalous currents occur for large fields if $%
\mu _{n}B/\Delta _{n}>\pi $ for some of the channels. We consider only the
linear regime in which the current cancels for a phase $\varphi \!=\!\pi
I(0)\delta _{+}\delta _{-}/ev_{F}(\delta _{+}+\delta _{-})$. This $\varphi -$%
junctions in a ring geometry generates a spontaneous current with a fraction
of a vortex threading the ring.
Let us point out that the Zeeman field couples to the 
momentum through the SO-coupling, acting as a gauge field that  
generate a $\varphi$-junction. Yet, this kind of symmetry argument is 
not sufficient, and the adiabatic QPC here plays an essential role in 
filtering and coherently mixing very few transverse channels. Indeed, 
no such effects have been obtained in wide junctions \cite{bezuglyi} and
quantum dots \cite{dellanna}. An exception is Ref. [\onlinecite{krive}] were a 1D case was considered. 

As in the zero-field case, a small change in $V_{g}$ shifts the resonance from
 $E_F$ and the current becomes a smooth function of the phase,
characteristic of a non-resonant junction. Remarkably, with more than one
pair of transmitting channels the CPR presents new effects, displayed 
for instance in the second conductance plateau, Fig.
3(c) and 3(d). In Fig. 3(c), we show a value of $V_{g}$ for which two
resonances---with different values of their parameters $\mu _{n}$ and
$\delta_{n}$---lie at $E_F$.

Changing $V_{g}$ shifts each resonance by a different
amount. The total CPR now results from the superposition of contributions with
different step-like structures and different $\varphi$-shifts. 
\textit{This leads to a
critical current $I_{c}$ that depends on the current direction}. We define the
critical current asymmetry as $I_{c}^{+}/I_{c}^{-}$ where $I_{c}^{+}$ and
$I_{c}^{-}$ are the critical currents for each flow direction. Figure 4 shows
that for physical values of the parameters a large asymmetry can be obtained by
tuning the gate voltage. The magnitude of the asymmetry depends on the detailed
structure of the dispersive Andreev states, which is determined by 
the interplay
between the SO-coupling, the external field and the QPC potential (see figs.
4(c) and 4(d)). We found that the asymmetry
$I_{c}^{+}/I_{c}^{-}$ can be larger than 3. These large assymmetry values are the main result of our work.

\begin{figure}[t]
\includegraphics[width=.45\textwidth]{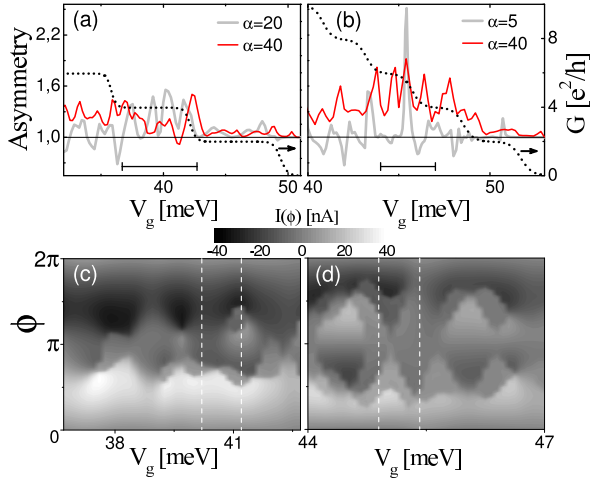}
\caption{Critical current asymmetry vs. $V_g$ for QPC1 (a) and QPC2 (b) and
different values of $\alpha$ (in meVnm) with the same applied field of Fig. 3.
The conductance of the QPCs is also shown (dotted line). The lower panels are
maps of the supercurrent for QPC1 with $\alpha\!=\!20$ (c) and QPC2 
with $\alpha\!=\!5$
(d) in the [$\phi$,$V_g$] plane. The $V_g$-scales of these maps are shown with
horizontal bars in panels (a) and (b). The vertical dashed lines correspond to
the curves of Fig. 3(c) and 3(d). } \label{FIG4}
\end{figure}

In summary, we have shown that a spin-polarizing QPC brings new physics to the JJs. 
While the most relevant effect is the new mechanism to generate critical current asymmetries, the 
device shows other interesting properties that highlight the effects of the SO-coupling on the Andreev states: \textit{i}) the supercurrent generates a 
magnetization in the 2DEG, being larger at the core of the QPC; \textit{ii}) an 
external in-plane magnetic field induces an anomalous current at zero phase 
difference (a $\varphi$-junction) \cite{krive}. In the latter case, we obtain
$\pi$-junctions for some values of the 
gate voltage (as for $\alpha\!=\!0$),  while in general 
$\varphi<\pi$. Such junctions, tunable both by an external flux and a 
Zeeman field, may have applications in SQUIDs or superconducting 
quantum bits.
With more than one transmitting channel, the external field induces a
large critical current asymmetry if the QPC potential $V_g$ is properly tuned.
Even for moderate values of the SO-coupling, and realistic values of the
external field ($B<1T$), the asymmetry can be quite large. The QPC is a central ingredient as it allows the control of the number and properties of the transmitting channels. 
These junctions act as supercurrent rectifiers in the interval
$\min(I_{c}^{+},I_{c}^{-})<|I|<\max(I_{c}^{+},I_{c}^{-})$, which can be controlled by adjusting the gate voltage. As this effect relies on the control of the spin polarization of the Andreev states, it generates a new alternative for supercurrent rectifiers based on pure spintronic effects. 

We acknowledge financial support through the ECOS-SECyT collaboration program
A06E03, ANPCyT Grants No 13829 and 13476 and CONICET PIP 5254. AAR and GU
acknowledge support from CONICET. DF and MA acknowledge support from 
ANR 
PNano Grant 050-S2.

\end{document}